# A framework for disentangling spatial and visual neural representations

**Mai M. Morimoto[1,2][*], Julien Fournier[3][*] and Aman B. Saleem[1][†]**

1. Department of Experimental Psychology, Institute of Behavioural Neuroscience, University College London, London, UK
2. Department of Life Sciences, Imperial College London, London, UK (current address)
3. Sorbonne université, CNRS, INSERM, Centre de Neuroscience NeuroSU, F-75005 Paris, France

[*] Contributed equally, [†]Corresponding author

## Abstract

Neurons in cortical areas often integrate signals from different origins. In the primary visual cortex (V1), neural responses are modulated by non-visual context such as the animal's position. However, the spatial profile of these position signals across the environment remains unknown. Here, we propose a new framework to disentangle visual and spatial contributions in virtual reality. This method relies on two principles: 1) a virtual corridor design that decorrelates vision and space through targeted cue repetitions and manipulations and 2) a Generalized Linear Model (GLM) that explicitly estimates visual contributions in retinotopic rather than environmental coordinates. In simulations, we demonstrate that this framework is highly specific (recovering spatial modulation only when present) and effectively captures the profile and weight of spatial gain fields across the environment. When applied to V1 recordings from mice navigating the virtual corridor, the model isolated significant spatial components in a substantial fraction of V1 neurons. The recovered spatial components exhibited heterogeneous, often multi-peaked, profiles. Application of this framework to large-scale recordings may provide a robust approach to characterize the nature of spatial signals modulating sensory processing across brain areas.

## Highlights

- We propose a quantitative framework combining virtual reality and a GLM model to disentangle visual and spatial neural representations
- Applying this framework on synthetic data showed low false positive rates for detection of spatial components
- Spatial and visual components of mouse V1 single unit neural responses could be disentangled and isolated
- This framework can be used as a discovery tool to characterise visual and spatial representations across brain regions

## Introduction

Disentangling the visual and spatial contributions to neural responses is challenging because visual cues are typically predictive of the animal's location, creating a tight coupling between vision and space. Virtual Reality (VR) provides a way to partially break this coupling by allowing precise control over visual cues associated with different locations. Previous work has investigated spatial modulation in the mouse visual cortex by using strictly repeating VR environments and inferring spatial signals from the difference between the neuron's peak response and its response at another visually identical location (Saleem et al. 2018; Diamanti et al. 2021). While this approach clearly demonstrated the existence of spatial modulation in V1, measuring it only at the neuron's peak response left two questions unresolved: what is the profile of V1 spatial signals across the entire corridor and what is the weight of this contribution relative to the overall response?

To address these questions, we developed a framework combining two key components: a vision model that explicitly accounts for visual receptive fields and a VR design that further breaks the tight correlation between vision and space. We designed a virtual environment with multiple out-of-phase repetitions of visual cues, and also included targeted manipulations, where landmarks were occasionally swapped or omitted. These manipulations provided additional information for separating visual and spatial contributions. In parallel, we implemented a GLM model estimating visual responses in retinotopic rather than environmental coordinates to capture what each neuron can realistically sample visually. Together, this combination of targeted VR design and constrained modelling provided a framework to disentangle spatial and visual components of V1 responses across the VR environment.

We first validated this framework using simulated neurons and confirmed that the model could recover the profile and the relative weight of spatial signals while maintaining high specificity (false positive rate <1%). We then applied this framework to a dataset of electrophysiological recordings from V1 in mice navigating the VR corridor. Within this dataset, the model identified significant spatial modulation in a substantial fraction of neurons. These spatial contributions exhibited diverse profiles - while some showed single peaks, many exhibited multiple peaks distributed across the corridor. Our GLM model also provided estimates of visual kernels, which revealed that some neurons were consistently suppressed by salient landmarks. In addition, the model included an "omission" component, designed to capture responses selective to the omission of a landmark (Fiser et al. 2016; Furutachi et al. 2024). In the V1 dataset analysed here, neurons showing significant omission-specific modulation were rare. Future applications of this framework may help reveal how spatial representations are structured and organized across brain regions and environments.

## Results
### A framework to disentangle vision and spatial components

To disentangle visual and spatial components, we designed a virtual reality (VR) corridor containing repeated landmarks and background textures (Figure **1A** and **1B**). Similar to previous works (Saleem et al. 2018; Diamanti et al. 2021), two landmarks were used: a vertical grating ($L_1$) and a plaid ($L_2$) each appearing twice along the 200 cm corridor, with 40 cm spacing between them. While the landmark repetition could in principle be sufficient to separate visual from spatial contributions, we further decoupled these variables by introducing manipulations on a subset of trials: in 8% of trials the landmarks at 80 and 120 cm were swapped, and in another 16% of trials, one of these two landmarks was skipped (Figure **1B**). Additionally, the background consisted of a smoothly varying white noise texture that repeated with a spatial period distinct from the landmark spacing (e.g. 52 cm). This design not only reduced the correlation between visual cues but also ensured that when a landmark was skipped, the exposed background segment was presented at other positions in the corridor, allowing its visual response to be estimated independently.

While this VR design reduces correlations between vision and space, it is not sufficient on its own to reliably estimate spatial modulation across all positions. Even with landmark manipulations, every position in the corridor is still characterized by a unique combination of landmarks and background textures if the visual scene is integrated globally. However, V1 neurons integrate the visual scene locally: they have receptive fields that typically do not extend beyond 60-80° in width (Niell and Stryker 2008; Bonin et al. 2011; Ayaz et al. 2013). We reasoned that this RF constraint could be exploited to further disentangle vision and space. We therefore designed a GLM in which the visual component is anchored in retinotopic space, allowing the model to capture what each neuron realistically samples by optimizing its RF position (Figure **1C** and **1D**). Specifically, visual space was binned in 5° steps, and for each neuron an 80° RF window was shifted across the hemifield (0°-120°) to determine its optimal position (Figure **1D**).

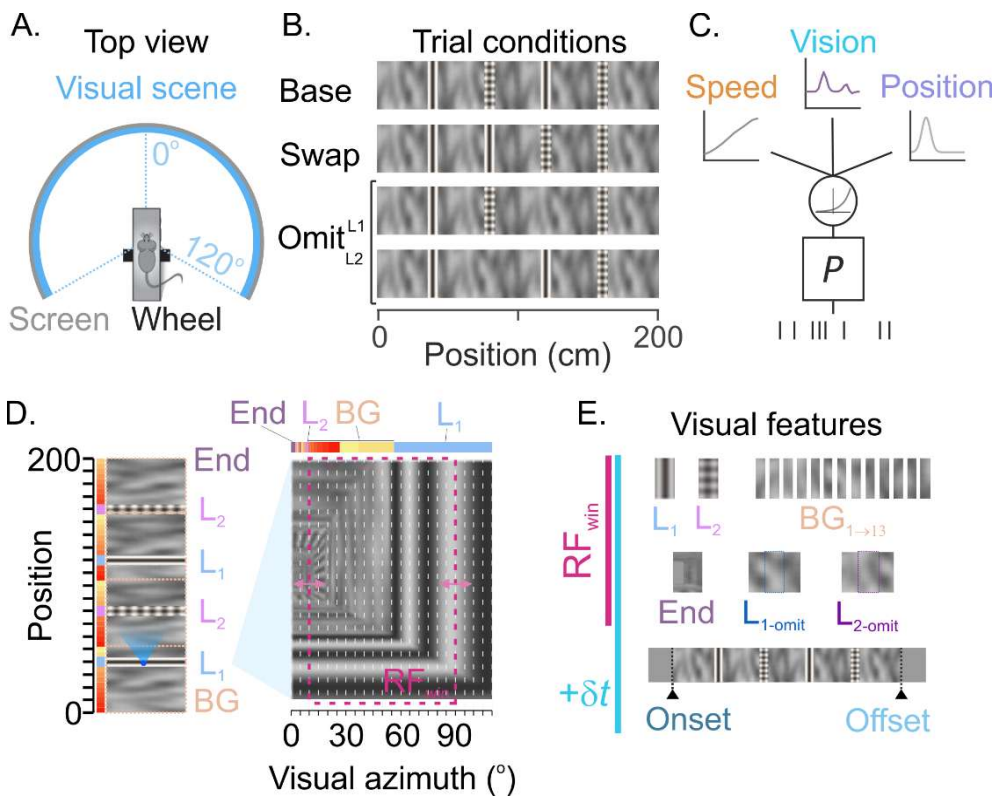

**Figure 1. Experimental and GLM framework for disentangling visual and spatial signals. A.** Top-down view of the virtual reality setup. Head-restrained mice run on a polystyrene wheel to navigate a virtual corridor projected onto a spherical screen spanning 240° of the visual field (described in Muzzu et al. 2021). **B.** Trial conditions. In base trials (76%), the mouse traversed a 200cm-corridor containing alternating landmarks spaced by 40 cm (a vertical grating $L_1$, and a plaid $L_2$), superimposed on a background texture repeating with a distinct spatial periodicity (e.g. 52 cm). In Swap trials (8%), the second and third landmarks were exchanged. In Omit trials (16%), one of these two landmarks were omitted, revealing the underlying background texture. **C.** Schematic of the Generalized Linear Model (GLM). The model estimates the contributions of visual features, running speed and animals' position to the neural response, assuming contributions are combined through an exponential nonlinearity to generate firing rates through a Poisson process (P). **D.** Transformation of visual cue positions from corridor to retinotopic coordinates. *Left*, Top-down view of the corridor showing landmarks locations and repeating background segments (coloured bars). *Right*, Visual textures projected into retinotopic coordinates (coloured bars). White dashed lines delimit the 5° bins used to discretize the hemifield. The purple dashed rectangle represents the 80° receptive field window, shifted across the entire hemifield (arrows) to identify the optimal RF location for each neuron. **E.** Visual predictors used in the model. Predictors include the two landmarks ($L_1$, $L_2$), the background segments (BG 1-13), the corridor's end wall, and omit-specific events ($L1_{omit}$, $L2_{omit}$), all estimated within the optimized RF window. VR onset/offset responses were modelled in the time domain. All visual predictors were time-shifted by a visual latency parameter ($\delta t$) optimized for each neuron.

Visual responses are also inherently delayed relative to the stimulus and visual and spatial signals most likely operate with different latencies. Because the animal runs with variable speeds from trial to trial, this temporal discrepancy can be leveraged to further separate visual and spatial contributions. Therefore, the GLM model also included a visual latency parameter which was optimized for each neuron and applied consistently to all visual predictors.

Visual predictors consisted of landmarks $L_1$ and $L_2$, background segments, the end wall of the corridor, and omission-specific components (Figure **1E**). Visual responses to onset and offset responses at trial start and end were modelled in the time domain, with the same latency shift as other visual predictors.

Since V1 neurons are strongly modulated by running speed (Niell & Stryker, 2011; Saleem et al., 2013), the model also incorporated running speed as a predictor. Together these components defined the default model (Vision + Speed), against which we tested the significance of adding the animals' position as a predictor (Vision + Speed + Position). We performed this comparison on held-out data using a likelihood ratio test (Zhang et al. 2023; Hardcastle et al. 2017; Acharya et al. 2016). This nested procedure ensured that the contribution of the position kernel was not overestimated.

Linear outputs of all components were summed and converted to firing rates with an exponential function. Spike times were obtained by a nonhomogeneous Poisson process driven by the instantaneous firing rates. Note that although linear in construction, this formulation is mathematically equivalent to a multiplicative model where the exponentiated outputs of each kernel combine multiplicatively in the spiking output; therefore, in the following, estimated kernels will be visualized as multiplicative gains by taking the exponential of their linear form (Pillow et al. 2008).

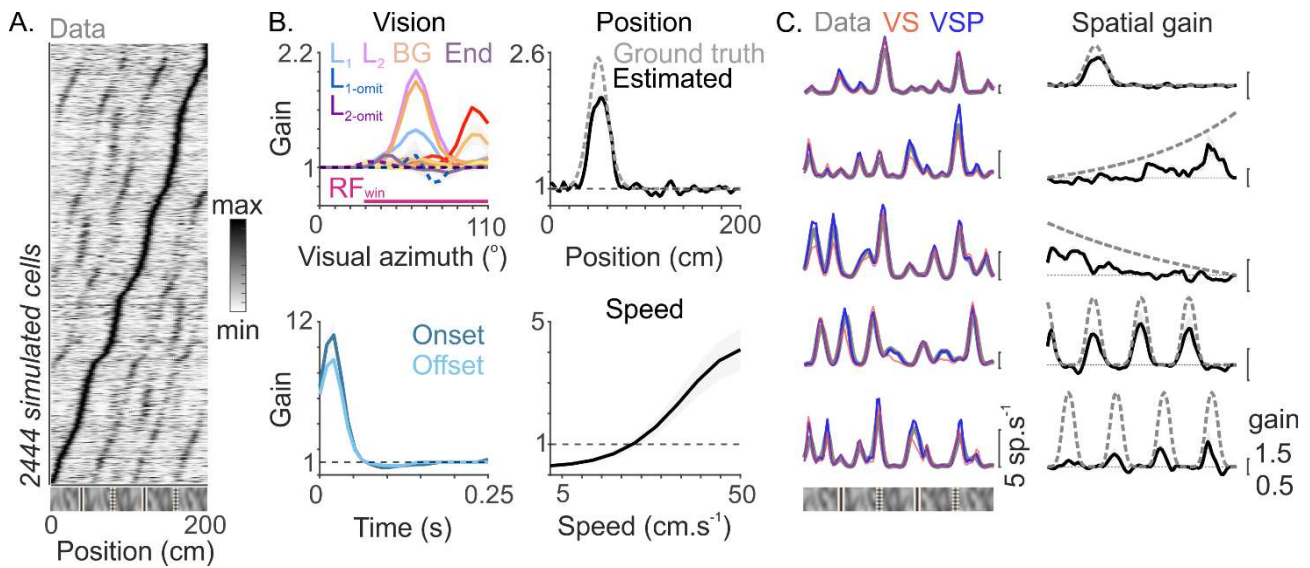

**Figure 2. Visual and spatial components estimated from simulated neurons A.** Normalized firing rate maps for the entire population of simulated neurons (n = 2444), computed from the simulated firing rates (*Data*) and sorted by the position of their peak response. **B.** Model components estimated for a simulated neuron. *Left*, Visual kernels (Landmarks, background, end-of-corridor and omission) estimated in retinotopic coordinates within the optimal RF window (pink bar) and VR onset/offset kernels estimated in the time domain. *Right*, Spatial (top) and speed (bottom) kernels. The grey dashed line indicates the ground-truth spatial profile used in the simulation. **C.** Firing rate maps and spatial kernels (spatial gain) for five example neurons simulated with distinct profile of spatial modulation. *Left*, Firing rate maps computed from the simulated spike train (Data, grey) or from the best GLM model, with (VSP, blue) or without (VS, red) the spatial component. *Right*, Recovered spatial kernels overlaid on the ground truth profiles for the same neurons.

**Separation of visual and spatial components in synthetic data**

The GLM model successfully disentangled visual from spatial components in synthetic data and importantly, was highly specific: spatial modulation was retrieved only when present, with a very low rate of false positives (Figure **2** and **3**).

To validate our framework, we generated synthetic neural data with known response components, RF position and visual latency and asked whether the GLM could recover spatial modulation across the corridor. Half of the synthetic neurons were assigned a spatial component, implemented as a Gaussian-, grid- or ramp-shaped profile; the rest of the neurons had only visual and speed components. The resulting firing rate across the corridor qualitatively resembled those observed in real V1 recordings (Figure **2A**). We then applied our GLM to estimate visual, speed and spatial components, testing for each neuron whether adding position significantly improved prediction (Figure **2B**, **2C** and **3A**). Neurons were considered as significantly modulated by spatial positions when they met two criteria: 1) the full model (Vision + Speed + Position, VSP) significantly outperformed the default Vision + Speed model (VS; likelihood ratio test on held-out data, $p < 0.05$) and 2) the weight of the spatial contribution (defined as the relative improvement in log likelihood when adding position to the model) was higher than 1%. This weight threshold was empirically defined from the ROC curve to ensure a false positive rate < 1% (Figure **3B**).

The model showed high specificity: spatial modulation was retrieved with a true positive rate of 48% and a false positive rate < 1%. The recovered spatial profiles closely matched the ground truth profiles, although amplitudes were often underestimated, particularly at positions overlapping with strong landmark responses (Figure **2C** and **3A**). This effect was more evident for grid and ramp-like profiles, where local fluctuations in amplitude clearly reflected an interaction with visual responses at landmark locations. Nevertheless, correlations between retrieved and ground truth profiles remained high (Figure **3A** and **3D**).

The weight of the retrieved spatial components correlated with the ground-truth weight measured from the simulated kernels, although it was generally underestimated (Figure **3C**). This bias further supports the specificity of our model, confirming that spatial contributions are not overestimated.

While the latency was accurately recovered in the vast majority of cases (Figure **3G**), the estimation of the RF position showed a systematic bias: the optimal RF window was generally estimated to be more lateral than frontal (Figure **3H**). Nevertheless, given the 80° width of the RF window, the model was still able to cover most of the RF for most neurons (Figure **3I**).

We explored the sensitivity of our framework to key model parameters. First, the temporal discrepancy between visual and spatial signals proved critical. In our model, since spatial signals are assumed with zero lag, the visual latency reflected this discrepancy entirely. We observed that the accuracy of the recovered spatial profiles depended heavily on the visual latency: for latencies below 100 ms, spatial estimates were more frequently contaminated by visual responses, whereas for longer latencies, the recovered profiles were more accurate (Figure **3E**). Second, the periodicity of the background texture influenced the model's sensitivity: when the background periodicity was increased to 104 cm (from 52 cm), the false positive rate remained low (~1.4%) but the true positive rate dropped from 48% to 16% (Figure **3F**). This reduction is likely due to the reduced number of texture repetitions combined with the larger number of predictors (~ x2). Finally, the inclusion of 'swap' and 'omission' trials in our experimental design proved important: when the model was fitted without landmark manipulations ('base trial only', Figure **3F**, right), the true positive rate decreased from 48% to 33% (under a spatial weight threshold guaranteeing a false positive rate below 1%).

**Profile and weight of spatial components in V1 data**

When applied to V1 recordings, our framework identified significant spatial modulation in a substantial fraction of neurons, with spatial profiles ranging from single peak to multi-peaked patterns (Figure **4**).

To ensure familiarity with the environment, mice were trained to run through the VR corridor (base condition with no manipulations) for 7–15 days before recordings. We then recorded neural activity from V1 using 32-channel silicon probes (Figure **4A**) while mice navigated the VR corridor with occasional landmark manipulations. Across 15 sessions from 7 mice, we isolated ~700 units, of which 367 were included in further analysis based on showing significant modulation of firing across the corridor (Figure **4B**) and a default GLM (Vision + Speed) that significantly outperformed a constant firing-rate model. This dataset included sessions with both short (52 cm; 7/15 sessions) and long (104 cm; 8 / 15 sessions) background periodicities.

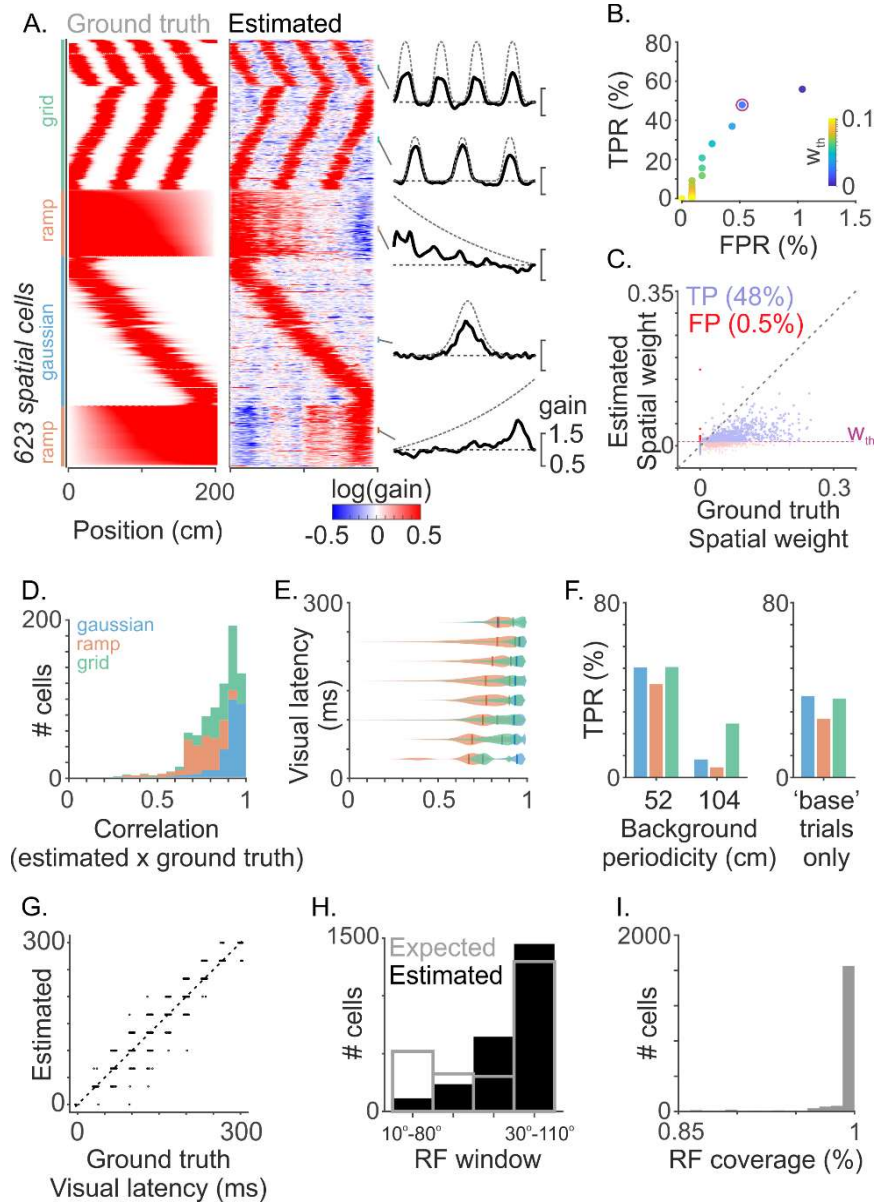

**Figure 3. Model performance on simulated data. A.** *Left*, Ground-truth and estimated spatial gain fields for the simulated neurons identified as significantly modulated by position (n=623). Neurons are ordered by the similarity of their ground-truth profiles, revealing the different shapes used in our simulations (Gaussian, ramp or grid-like patterns). *Right*, Examples of spatial profiles averaged over four consecutive neurons with high similarity (Estimated: solid black; Ground-truth: dashed grey). **B.** Receiver Operating Characteristic (ROC) curve showing True (TPR) and False positive rates (FPR) as a function of the threshold used for the spatial weight ($w_{th}$). The purple circle indicates the selected operating point ($w_{th} > 1\%$) for which FPR < 1%. **C.** Comparison of estimated versus ground-truth spatial weights for all simulated neurons (Purple: True positive; Pink: True negative; Red: False positive; Gray: True negative. The estimated weight was defined as the fraction of improvement in log-likelihood when adding position to the model. The ground-truth weight was defined as 1 - r, where r is the Pearson's correlation between the simulated firing rate with and without the spatial contribution. The inclusion threshold on the spatial weight ($w_{th}$) is indicated by the dashed purple line. **D.** Distribution of Pearson's correlation between estimated and ground-truth spatial profiles, color-coded by the shape of the simulated profile (Gaussian, ramp or grid-like). **E.** Same as in D but stratified by the visual latency of the simulated neurons. **F.** *Left*, True positive rates (TPR) for different spatial patterns, compared between simulations using background textures of different periodicities. All other analyses performed on simulations are based on a background with periodicity of 52 cm. *Right*, True positive rates (TPR) for simulations with no 'swap' or 'omission' trials in the experimental design ('base' trial only). **G.** Comparison of the visual latency estimated by the model with the ground-truth latency. **H.** Distribution of the centre of the optimal RF window recovered by the model (black), compared to the expected distribution based on the peak positions of the ground-truth RFs (grey). **I.** Distribution of the fraction of the ground-truth RF envelope covered by the optimal 80° RF window estimated by the model.

The distribution of neural responses along the corridor was consistent with previous studies (Saleem et al. 2018; Diamanti et al. 2021) (Figure **4B**). While peaks tiled the entire corridor, they clustered more frequently around landmarks, suggesting a strong visual component in those units (Figure **4B**). To disentangle visual and spatial contributions, we applied the GLM model to this dataset, first estimating visual and speed components for each unit, and then testing whether adding position to the model significantly improved predictions (Figure **4C** and **4D**).

In this dataset, our method identified ~20% (76/367) of neurons with a significant spatial modulation. As demonstrated in our simulations, the recovery rate of our framework is sensitive to the periodicity of the background texture. To contextualize this proportion, we ran simulations matching the proportion of background periodicities used in the experimental data and found a true positive rate of ~31%. This indicates that the 20% observed in V1 recordings represents a lower bound; adjusting for the sensitivity of our method, the true prevalence of spatially modulated neurons in V1 could be up to ~3 times higher than detected.

Comparison of the model output with or without spatial contributions (VSP versus VS) showed that spatial modulation could either suppress or boost responses at specific locations (Figure **4C**). The profiles of these spatial components were heterogeneous: ordering them by similarity showed that some peaked around the beginning or end of the corridor, while many exhibited multiple peaks distributed across various positions (Figure **4E**). While the recovered spatial profiles sometimes coincided with the peak of the overall response, this overlap was not systematic (Figure **4F**). On average, the spatial component accounted for ~7% of the explainable response (Figure **4G**).

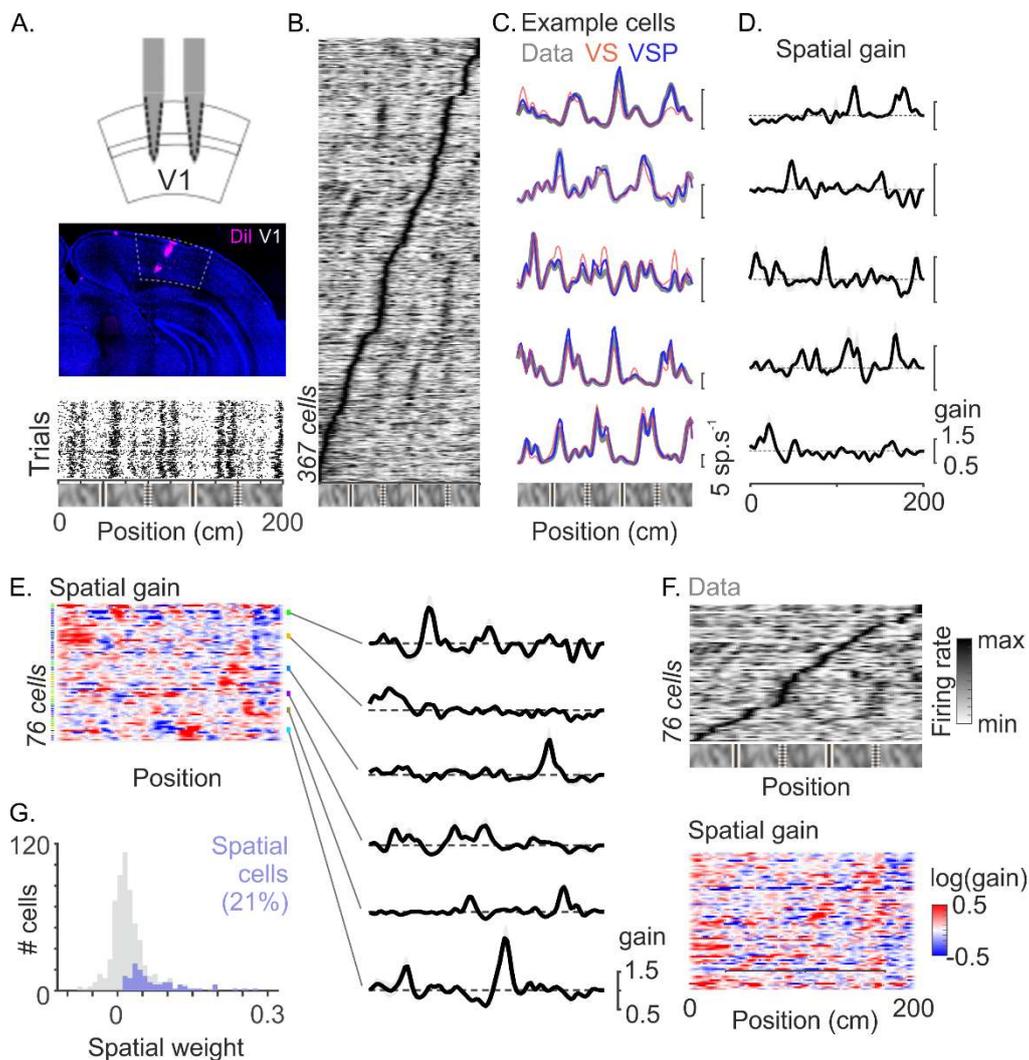

**Figure 4. Recovering spatial components from V1 neural responses. A.** *Top*, Schematic of dual-shank 32-channel silicon probe recording from mouse V1. Electrode placement was confirmed histologically from DiI deposit along the electrode track. *Bottom*, Raster plot of spikes from an example neuron during "base" trials. **B.** Normalized firing rate maps for the entire population of analysed V1 neurons (n = 367), sorted by the position of their peak response along the track. **C.** Firing rate maps of four example neurons recorded simultaneously and identified as having a significant spatial modulation. Maps were computed from the actual recorded spike trains (Data, grey) and from the best model with (VSP, blue) or without (VS, red) including the spatial component in the model's prediction. **D.** Estimated spatial gain fields for the same neurons as shown in C. **E.** *Left*, Spatial gain fields for all significantly modulated neurons (76/367, ~21%), sorted by profile similarity. Coloured ticks on the left indicate the recording session/animal for each unit. *Right*, Examples of spatial profiles averaged over clusters of four neurons with high similarity. **F.** Firing rate maps (*top*) and corresponding spatial gain fields (*bottom*) for the neurons with significant spatial modulation, both sorted by the peak position of the firing rate maps. **G.** Distribution of the weight of the spatial contribution for neurons identified as significantly modulated by spatial positions (purple, mean: 0.074) and those that were not (grey, mean: 0.011).

Together, these results demonstrate that our framework can detect significant spatial modulation in V1 responses, and that its profile is heterogeneous, often manifesting as multi-peaked gain fields scattered across the corridor.

**Visual components of V1 responses in a virtual corridor**

In addition to isolating spatial components, our framework also provides estimates of purely visual responses, including those driven by landmarks ($L_1L_2$), background segments (BG), corridor end, VR onset/offset events and omission of landmarks (Figure **5**). In simulations, the kernels estimated in retinotopic space did not strictly match the Gaussian-shaped receptive fields used as ground truth (Figure **2A**) but once convolved with the visual scene and projected back to corridor coordinates (Figure **5A**), they qualitatively reproduced the ground truth profiles (Figure **5B**). While the amplitudes of responses to $L_1/L_2$ and background segments sometimes deviated from the ground truth, the overall profiles and dynamics were highly correlated (Figure **5C**).

In actual V1 recordings, landmark responses showed some diversity (Figure **5D**). As expected, some neurons preferred the grating ($L_1$), but more often preferred the plaid ($L_2$) or responded to both. Notably, some neurons exhibited systematic suppression at landmark locations (Figure **5D**). In these neurons, responses to background features were often dominant, shifting the overall peak response to between rather than at landmark locations. Responses to background features also exhibited diverse patterns: while some neurons responded regularly to several of the background segments, others responded selectively to only certain repetitions, with suppression by other background segments at intervening positions (Figure **5E**).

**Capturing landmark omission responses in V1**

V1 neurons can respond to the omission of an expected visual stimulus or the presentation of an unexpected stimulus (Fiser et al. 2016; Furutachi et al. 2024). In our protocol, the omission trials may trigger such responses: the absence of the expected landmark revealing an unexpected piece of the background texture. To account for responses that would be selective to this unexpected visual event, our GML model included an "omission-specific" component, expressed in retinotopic coordinates and only present during omission trials.

In simulations, omission-selective components were recovered with a rate of ~30% (27.4%, 255/932). This low recovery rate reflects a model trade-off: such responses can often be explained, at least partially, by other visual components, especially the background segment revealed by the absence of the occluding landmark.

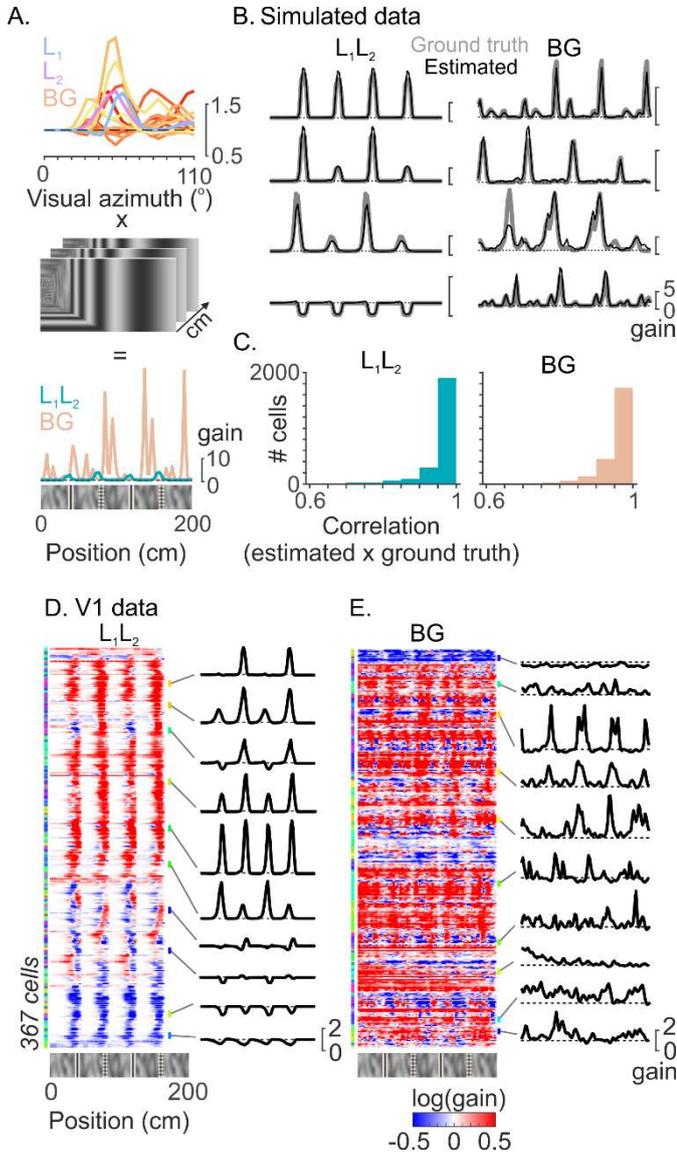

**Figure 5. Recovering visual components of V1 neural responses. A.** Projection of visual kernels into corridor's coordinates. Raw visual kernels (*top*; $L_1$, $L_2$, BG) estimated in retinotopic space were convolved with the layout of visual textures in each position (*middle*) to generate the effective visual profile as a function of corridor position (*bottom*). **B.** Examples of visual components recovered in simulations for landmarks (*left*, $L_1L_2$) and background (*right*, BG) textures. Black lines indicate the model estimates; grey line, the ground-truth profile expected from the kernels used in the simulations. **C.** Distribution of Pearson's correlations between estimated and ground-truth response profiles for landmarks (*left*, $L_1L_2$) and background (*right*, BG) textures across the population of simulated neurons. **D.** *Left*, Landmark response profiles ($L_1L_2$) estimated for all analyzed V1 neurons (n = 367), sorted by similarity. Colored ticks on the left indicate the recording session/animal for each neuron. *Right*, Examples of mean landmark response profiles averaged over clusters of four neurons with high similarity. **E.** Same as in D for background response profiles (BG) sorted by similarity.

In actual V1 recordings, we found that 5% of neurons (19/367) exhibited omission-specific responses. When accounting for the ~30% recovery rate observed in our simulations, this suggests that at most ~15% of neurons may be truly selective for the absence of an expected landmark in the V1 dataset analysed here.

## Discussion

To our knowledge, this is the first framework that allows for disentangling visual and spatial neural representations through a combination of constrained modelling and targeted manipulations of the visual environment. Extending previous work that established the existence of spatial modulation in V1 (Saleem et al. 2018; Diamanti et al. 2021), the current framework allows for estimation of the visual and spatial profiles across the entire environment. Future work applying similar approaches across brain regions could shed light on the structure of spatial representations across the brain. In the following, we discuss critical design trade-offs and potential optimizations for future implementations of this framework.

### Optimising VR design for vision and space separation

While our framework can effectively isolate spatial components in visual neurons, the sensitivity of the method is intrinsically related to the design of the virtual environment. In future iterations, several optimizations of the visual textures and landmark manipulations could further improve the recovery rate of spatial components.

First, the alignment between visual textures and model predictors could be improved. In our current implementation, the background texture was generated independently of the GLM spatial binning, which requires segmenting the background texture based on the minimal visible area imposed by occlusions of the background by the landmarks. Future designs could explicitly align the spatial frequency of the background texture with the model's resolution. Second, reducing the periodicity of the background textures and increasing the frequency of landmark occlusions would help create more visually unique combinations, thereby providing the model with better leverage to decorrelate the contributions from the different textures.

Finally, the protocol for landmark manipulation could be refined to reduce sampling bias. In the current version, swap and omission were restricted to the second and third visual landmarks. Therefore, the background patterns occluded by the first and fourth landmarks were exposed less often than other segments. Future iterations could ensure that manipulations are applied uniformly across all salient landmarks to guarantee equal sampling across all visual cues.

These optimizations must be balanced with one crucial constraint: the stability of the animal's visual and spatial representation. The trade-off in designing such a VR is to loosen the correlation between vision and space while preserving a stable cognitive representation. For instance, if salient visual cues are repeated too often, the animal may perceive a sequence of repeating corridors. Conversely, if landmark manipulations are too frequent, it could induce a remapping that could lead to interpreting a context shift rather than a local perturbation. Therefore, an optimal VR design should balance repetitions to allow the model to separate vision from space, and saliency to induce a stable and continuous spatial representation. Here, in our VR design, each manipulation type is 8% of total trials during which we assume that the spatial representation stays constant.

Finally, the choice of the best textures and VR structure will likely depend on recording area and species, and a one fits all solution may not exist. Here we have built on previous work in V1, and opted for landmark and background texture combinations, but it is possible that these distinctions may not be required in other contexts. Applying this framework to other brain regions or species will require designing a visual environment that drives the local population maximally while respecting the specific cognitive constraints of the animal.

### Choice of GLM kernels and modelling approach

The predictors included in our default GLM model were selected based on established properties of visual RFs (Hubel and Wiesel 1962; Niell and Stryker 2008) and running speed modulation in V1 neurons (Saleem et al. 2013; Horrocks et al. 2024). However, this baseline description could be refined to incorporate more complex features. For instance, V1 receptive fields often exhibit significant nonlinearities such as centre surround suppression (Sillito and Jones 1996; Polat et al. 1998; Ayaz et al. 2013). Future work could evaluate the impact of these nonlinearities on the model's sensitivity and if necessary incorporate specific gain control mechanisms to account for them. Additionally, while our RF estimates were validated through simulations, future experiments could use RFs mapped independently with standard mapping protocols (Smith and Häusser 2010; Bonin et al. 2011).

Furthermore, while eye movements in head-fixed mice navigating linear corridors are usually limited (Meyer et al. 2020), small compensatory drifts can still shift the retinotopic frame. Incorporating eye-tracking data as an additional covariate to adjust the RF window on the animal's gaze would likely help recover more accurate visual components and reduce the residual variance.

In terms of model architecture, our framework can easily be adapted to the question at stake. Here, we defined the visual model (Vision + Speed) as the null hypothesis, to be strictly conservative about the existence of spatial modulation. However, the model is agnostic to this hierarchy and this logic can be adapted or completely inverted. For instance, for hippocampal or entorhinal recordings, where spatial coding is dominant (O'Keefe and Dostrovsky 1971; Fyhn et al. 2004), one may define a "spatial" model (Position + Speed) as the null hypothesis and test for the significance of purely visual contributions (Purandare et al. 2022).

Our GLM assumes a Poisson spiking process driven by an exponential nonlinearity. Consequently, while the model sums inputs linearly, the exponential link implies that visual and spatial signals interact multiplicatively in the spiking output. This formulation aligns well with how multisensory or sensorimotor information is thought to be integrated within cortical areas (Andersen and Mountcastle 1983; Pouget et al. 2002; Chang et al. 2009). However, other interactions (multiplicative at the input stage or purely additive) may be valid. Future exploration may thus compare these formulations and determine which operation best captures the integration of spatial signals in V1 neurons.

Finally, in our implementation, we prioritized specificity over sensitivity. While the statistical procedure to test for significance could be adjusted to increase sensitivity, we opted for a conservative approach to ensure that the model detected spatial modulation only when present. As a consequence, the proportions of spatially modulated or omission selective neurons reported here should be interpreted as lower bound estimates of their true prevalence in the overall population.

## Biological insights and future directions

In the present study, our framework revealed that a substantial fraction of V1 neurons are significantly modulated by spatial position. The profiles of these spatial gain fields were highly heterogeneous, often showing muti-peaked profiles rather than single subfields. Given the size of the current dataset, it is difficult to definitely categorize these profiles into distinct functional classes such as place, grid or ramp-like profiles. Consequently, whether V1 spatial modulation is dominated by specific spatial patterns remains an open question. To resolve this, future work could apply this framework to recordings of larger scale, and potentially with cell type specificity. Additionally, extending this analysis to regions projecting to V1 such as higher visual areas or the retrosplenial cortex (Morimoto et al. 2021; Yao et al. 2023) could provide insights into whether V1 inherits those complex spatial profiles or computes them locally.

The dataset analysed here also provided insights into omission specific responses. In our specific experimental setting, responses to the absence of an expected landmark were rare (~5%). However, because omission trials constituted only 8% of the trials (per landmark), our method had limited sensitivity for this component (recovery rate ~30%). Therefore, the prevalence of omission selective neurons in our environmental settings remains an open question requiring further investigation with higher statistical power.

Finally, another area for future research is the temporal evolution of V1 spatial signals. It remains to be seen how spatial patterns emerge during learning, how they stabilize as the animal familiarizes itself with the environment and whether or not they remap across sessions. Future work may use the proposed framework to track these dynamics, offering a path to understand how vision and space are integrated over the course of experience.

## Acknowledgements
This research was supported by The Sir Henry Dale Fellowship from the Wellcome Trust and Royal Society (200501), Biotechnology and Biological Sciences Research Council (BBSRC; BB/R004765/1), UKRI Frontier Research Grant (EU underwrite; EP/Y024656/1), and by the Human Frontier Science Program (RGY0076/2018) to AS, Japan Society for the Promotion of Science (JSPS) Overseas Research Fellowship and the Imperial College Research Fellowship to MM and the Human Frontier

**Contributions**

Conceptualization: MM, JF and AS; methodology: MM, JF, AS; experimentation: MM; investigation and analysis: MM and JF; validation: MM, JF and AS; visualization: MM and JF; writing first draft: MM and JF; revision, review, editing: MM, JF and AS; funding coordination, project administration and resources: AS.

**Data and code availability**

Data collected in this study is available upon request. Any additional information required to analyse the data reported in this paper is available from the lead contact upon request.

## Methods

### Experimental model and subject details

All experimental procedures were conducted in accordance with the UK Animals Scientific Procedures Act (1986). Experiments were performed at University College London under personal and project licenses released by the Home Office following appropriate ethics review. Data were collected from 7 mice (C57BL6 wild-type, 3 females and 4 males, aged 10-28 weeks). They were housed in groups of maximum five under a 12-hour light/dark cycle, with free access to food and water.

### Surgical procedure

Mice were implanted with a custom-built stainless-steel metal plate on the skull under isoflurane anaesthesia and allowed to recover with analgesia (Carpofen). The implant was positioned to leave the area above the right visual cortex accessible for recordings. After surgery, mice were singly housed.

Seven days following the surgery, all but one mouse underwent an habituation period, consisting of daily sessions to acclimatize them to head-fixation and VR environment. After this habituation, a small craniotomy was performed over V1 (2.5 mm lateral, 0.5 mm anterior to lambda). The dura was left intact to preserve the brain tissue and maximize recording stability. First recordings started 4-24 hours following the craniotomy.

Multiple recording sessions were performed on each animal (one per day), for a total of 17 recording sessions included in this dataset. All electrophysiological recordings were carried out during the dark phase of the animal's light cycle. Between sessions, the craniotomy was protected with a silicon cap (Kwikcast, World Precision Instruments).

### Virtual reality setup and behavioural protocol

The virtual reality setup was described previously (Muzzu and Saleem 2021; Zucca et al. 2025). In brief, mice were head-restrained and free to run on a polystyrene wheel (radius 10 cm), with their heads positioned in the geometric centre of a truncated spherical screen onto which the visual stimulus was projected. Visual stimuli were produced by reflecting the light of a projector (Casio Green Slim XJ-A257-UJ DLP; 60 Hz refresh rate), onto the internal surface of the dome via an hemispherical mirror; the projected VR image spanned 240° azimuth and 120° elevation (from -30° and 90°) with mean luminance ca. 10 cd.m². Visual stimuli were produced by custom MATLAB code as previously described (Saleem et al. 2018). Movement of the wheel was sensed with a rotary encoder (2400 pulses/rotation, Kübler, Germany), the output of which was copied to both the OpenEphys (Siegle et al. 2017) acquisition board and to an Arduino connected to the stimulus computer, so as to allow locomotion-dependent updating of the visual scene. A pseudo-random synchronising signal was sent to both the OpenEphys board and the stimulus computer.

Animals were habituated to the virtual environment for 7-15 days (one session per day) before recordings started. Animals did not receive rewards during any of the habituation or recording sessions. On each habituation day the animal was placed in the virtual reality apparatus for approx. 10 mins, during which it moved within the virtual environment by running on the treadmill.

The virtual environment was a linear corridor with 4 alternating landmarks (a gratings and a plaid repeating 80 cm apart) on a textured background (gaussian filtered white noise) similar to the environment described previously (Saleem et al. 2018; Diamanti et al. 2021). The total length of the corridor was 200cm in length, and the plaids and gratings were 8cm in width. The textured background repeated with a periodicity of 52 cm (n = 7/15 sessions), or 104 cm (n = 8/15 sessions). All textures were repeated on all walls of the corridor, covering the entire visual field of the animal.

During habitation, the animals experienced only the "base condition" which had no landmark manipulation. On recording days, we presented corridors with manipulated landmarks on 24% of the trials (swapping or omitting the second or third landmark) interleaved with the base condition in pseudo random order.

### Electrophysiological recordings and data processing

Extracellular activity was recorded using multi electrode silicon probes with two shanks of 16 channels each (ASSY-37 E-1, Cambridge Neurotech Ltd, Cambridge, UK) connected to an OpenEphys acquisition board sampling at 30 kHz (Siegle et al. 2017). Anatomical location of the recording site was confirmed via post-hoc histology targeting the DiI deposits left by the probe coating.

Spikes were detected and sorted using Kilosort 2 (Pachitariu et al. 2016), followed by manual curation using Phy to check cluster quality and remove noise artifacts. Spike trains were then binned at the VR frame rate (60 Hz, ~ 16.7 ms) for analysis.

Across all 15 recording sessions, ~700 units were isolated, out of which 367 were included in our analyses. Two criteria were used for inclusion: 1) significant modulation of the neuron's firing rate along the corridor and 2) a "default" GLM model (VS, Vision + Speed) that significantly outperformed a constant firing rate model (see below). Only recording sections during locomotion (animal run speed>1 cm.s$^{-1}$) were used for analysis.

### Generation of synthetic data

Each simulated neuron was assigned a Gaussian-shaped receptive field (RF) in retinotopic space (5° to 10° s.d.), with selectivity for the two landmarks ($L_1$, $L_2$), the end-wall of the corridor, a random subset of background segments and the VR onset and offset. RF peak positions were sampled uniformly between 10° and 120° of visual azimuth, and response latencies were drawn from a normal distribution centred at 150 ms, consistent with V1 peak response delays. The relative selectivity to each visual feature was chosen randomly, with the constraint that background components were capped at 150% of the peak response to landmarks. Additionally, for half of the neurons, we simulated selectivity to landmark omissions (at 80 or 120 cm) by including an additional visual component with the same RF parameters and an amplitude set to 20-40% of the neuron's maximal response to visual cues.

Half of the simulated neurons were assigned a spatial component. Spatial profiles were drawn from one of three shapes: unimodal (Gaussian), periodic (grid-like), or monotonic (ramp-like). The amplitude of these spatial components was set to 20-40% of the maximal visual response. Speed selectivity was simulated using tuning curves measured from real V1 data.

Finally, the linear outputs of all components (Visual, Speed and Spatial) were summed, passed through an exponential function and scaled to match the distribution of firing rates observed in our V1 recordings. The resulting instantaneous firing rate drove a Poisson process to generate discrete spike times, to which we subsequently applied the GLM estimation procedure.

### GLM framework and model fitting

To disentangle the contributions of visual, speed and position variables to neural responses, we modelled the spike train of each neuron using a Generalized Linear Model (GLM; Acharya et al. 2016; Hardcastle et al. 2017; Zhang et al. 2023).

Visual predictors included the two landmarks ($L_1$ and $L_2$), background segments (discretized into 4 cm intervals), the end-wall of the corridor, VR onset and offset, and specific predictors for landmark omissions (separate predictors for $L_1$ and $L_2$ absence).

Predictors for VR onset and offset were modelled in the time domain over a 250 ms window discretized in ~17 ms bins (one VR frame). All other visual predictors were defined in retinotopic coordinates by mapping corridor positions to visual angles $\theta$ using the transformation:

$$\theta = 90 - arctan\left(\frac{p}{w/2}\right) \times \frac{180}{\pi}$$

Where $p$ is the position in corridor's coordinates and $w$ is the width of the corridor. The visual hemifield (120° wide) contralateral to the recording site was discretized (5° bins), each predictor thus indicating the presence of a specific visual feature in a given angular bin at every time step.

To further constrain the model to realistic RFs, we included two additional hyperparameters: a visual latency and an RF window. The RF window was defined as an 80° aperture, outside of which visual predictors were set to zero. During optimization, we explored visual latency ranging from 0 to 300 ms (in steps of ~33 ms or 2 VR frames) and RF window centres ranging from 40 to 70 (in 10° steps). Overall, the visual features were represented by 396 predictors in the model (for a background with a periodicity of 52 cm).

To account for the known modulation of V1 neurons by locomotion, the model also included running speed. Speeds were capped at 50 cm.s$^{-1}$ and discretized into 5 cm.s$^{-1}$ bins, creating 10 predictors indicating the animal's speed range at each time step.

Finally, spatial position was parametrized by binning the corridor into 2-cm intervals, yielding 100 spatial predictors. To mitigate collinearity between spatial positions and the last end-wall texture, spatial predictors within the final 6 cm were excluded; within this range, the angular position of the corridor's end (< 45°) is indeed highly predictive of the linear corridor position (differing by less than 5%).

This model was fitted to the entire recording (including trial and inter trial periods) using a Maximum Likelihood Estimate (MLE) and assuming a Poisson noise distribution. Specifically, the instantaneous firing rate was modelled as an exponential function of the linear sum of input covariates:

$$R(t) = e^{(k_0 + k_{vis} \times x_{vis} + k_{spd} \times x_{spd} + k_{pos} \times x_{pos})}$$

Where $k_{vis/spd/pos}$ represent the kernel weights for visual, speed and position contributions and $k_0$ is a baseline offset. To mitigate discretization artifacts, the raw estimates of the kernel weights were smoothed with a Gaussian kernel (s.d. of one bin size).

We evaluated model performance using a 10-fold cross-validation scheme: the model was trained on 90% of the trials and tested on the held-out 10%, using a circular permutation until all trials were predicted.

To prevent overfitting given the high dimensionality of the predictor space, we applied L1 regularization (Lasso). The optimal regularization parameter was selected by maximizing the log likelihood of the held out data. Finally, the optimal hyperparameters for the visual components (RF window position and visual latency) were determined using a grid search, selecting the combination that maximized the likelihood of the held out data.

**Model selection and components' weights**

To evaluate whether spatial components contributed significantly to neural responses, we adopted a nested model comparison approach.

We first fitted the "default" model (VS, Vision + Speed) and compared its predictive performance against a null model (constant firing rate). If the default model provided a higher log-likelihood of the held-out data than the null model, we then tested the significance of the spatial component by comparing the log-likelihood of the held-out data under the full model (VSP, Vision + Speed + Position) to that of the default model (likelihood ratio test, p < 0.05). The significance of the omission components was assessed similarly, by comparing the log-likelihood of the best model with and without the omission components.

This evaluation was performed across the entire grid of visual latencies and RF window positions. The best overall model was the one identified for the combination of hyperparameters that yielded the highest likelihood.

To quantify the relative contribution of spatial signals, we estimated a spatial weight based on the improvement in model prediction when adding in spatial predictors. We first defined the log-likelihood increase (*LLHi*) for a given model as the difference between its log-likelihood and that of the null model (constant firing rate) on held-out data:

$$LLHi_{model} = LLH_{model} - LLH_{null}$$

The weight of the spatial component ($w_{pos}$) was then calculated as the fraction of predictive power that was gained by adding spatial predictors to the default model:

$$w_{pos} = 1 - \left(\frac{LLHi_{VS}}{LLHi_{VSP}}\right)$$

Intuitively, this metric ranges from 0 (or negative values) if spatial predictors add no information beyond vision and speed, and to 1 if spatial predictors account for all the predictive improvement over the null model.

Neurons were considered as spatially modulated when they met two criteria: 1) the spatial component was found significant when comparing the full model to the default model (see above) and 2) the weight of the spatial contribution was higher than 1%. This weight threshold was defined from the Receiver Operating Characteristic (ROC) curve, to ensure a false positive rate < 1% (**Figure 3B**).

**Quantification of response profiles**

To visualize and quantify the modulation of firing rates along the corridor, we computed firing rate maps. Spike counts and occupancy were aggregated in 2-cm bins across the corridor and smoothed with a Gaussian (s.d. of one bin size). Firing rate maps were then obtained by dividing smoothed spike counts by occupancy. Similar maps were computed for the firing rates predicted by the GLM model (without the spatial component), except no additional smoothing was applied since GLM kernels were already smoothed.

To assess the significance of firing rate modulation across the corridor, we used a similar approach to the GLM model selection. A neuron's firing rate was considered significantly modulated along the corridor if the log-likelihood of the held-out data under the firing rate map model was significantly higher than that of a null model (constant firing rate; likelihood ratio test, p < 0.05).

To compare visual and spatial kernels estimated by the GLM in the same reference frame, the estimated visual kernels were projected back into corridor coordinates. This was achieved by convolving the retinotopic kernels with the specific sequence of visual cues encountered at each position along the corridor, thereby generating the expected visual profile of the neuron in spatial coordinates (Figure **5A**).

Finally, to visualize the kernel profiles across the population, spatial and visual kernels were sorted by similarity. We computed the Manhattan distance between all pairs of normalized profiles and performed hierarchical clustering with 'average' linkage. The final order was obtained using optimal leaf ordering to maximize the similarity between adjacent profiles.